\documentclass[twocolumn,showpacs,preprintnumbers]{revtex4}
\usepackage{amssymb}
\usepackage{amsfonts}
\usepackage{amsmath}
\usepackage{graphicx}
\usepackage{dcolumn}
\usepackage{bm}

\input{tcilatex}

\begin{document}

\title{Isothermal reentrant effect in a mesoscopic cylindrical structure of
a superconductor coated with a normal metal layer}
\author{G. A. Gogadze}
\email{gogadze@ilt.kharkov.ua}
\author{S. N. Dolya}
\affiliation{B. Verkin Institute for Low Temperature Physics and Engineering of the
National Academy of Sciences of Ukraine 47, Lenin Ave., Kharkov 61103,
Ukraine.}
\date{\today }

\begin{abstract}
The coherent phenomena in mesoscopic cylindrical normal metal (N) -
superconductor (S) structures have been investigated theoretically. The
magnetic moment (persistent current) of such a structure has been calculated
numerically and (approximately) analytically. It is shown that the current
in the N-layer corresponding to the free energy minimum is always
diamagnetic. As the field increases, the magnetic moment (current) exhibits
jumps at certain values of the trapped magnetic flux and the NS structure
changes to a state with smaller absolute value of the diamagnetic moment.
This occurs when the persistent current is unable to screen the external
field. The magnetic moment increase stepwise and the system changes into a
new stable state. The magnetic field penetrates into a larger volume of the
N-layer. The new state has smaller absolute value of the diamagnetic moment.
Experimentally, this is interpreted as the presence of a paramagnetic
addition in the system (paramagnetic reentrant effect). The results obtained
are in qualitative agreement with the experiments conducted by P. Visani, A.
C.Mota, and A. Pollini, Phys. Rev. Lett. \textbf{65}, 1514 (1990).
\end{abstract}

\maketitle

\section{Introduction}

The Meissner effect in a mesoscopic cylindrical structure consisting of a
superconductor coated with a pure normal-metal layer has interesting
features due to the coherent quantum effects in the normal metal. They are
observable when there is a good contact between S and N constituents. This
problem within the quasiclassical Eilenberger formalism was first
investigated theoretically by Zaikin \cite{1a}.

Recent advanced technologies of preparation of pure samples have enabled to
investigate the coherence properties of mesoscopic samples taking a proper
account of the proximity effect \cite{2a}. The samples were superconducting
Nb wires with a radius $R$ of tens of microns coated with a thin layer $d$
of high-purity Cu, Ag or Au. Mota and co-workers \cite{3a} detected
surprising behavior of the magnetic susceptibility $\chi $ of a cylindrical
NS structure (N and S are for the normal metal and the superconductor,
respectively) at very low temperatures ( \textrm{T} $<$ $100$ mK ) in an
external magnetic field parallel to the NS boundary. Most intriguingly, a
decrease in the sample temperature below a certain point $T_{r}$ (at a fixed
field) produced a paramagnetic reentrant effect: the decrease of magnetic
susceptibility of the structure is changed to an unexpected grow. A similar
behavior was observed in the isothermal reentrant effect in a field
decreasing to a certain value $\mathrm{H}_{r}$ below which the
susceptibility started to grow sharply. It is emphasized in Ref. \cite{4a}
that the detected magnetic response of the NS structure is similar to the
properties of the persistent currents in mesoscopic normal rings.

There have been numerous attempts to explain the paramagnetic reentrant
effect theoretically \cite{5a,6a,7a}. However, the predicted amplitudes of
the effect were too small ( except for \cite{6a} ) to account for the
experimental facts. Fauchere, Belzig and Blatter \cite{6a} explain the large
paramagnetic effect assuming strong repulsive electron-electron interaction
in noble metals. The proximity effect in the N metal induces an order
parameter will be shifted by $\pi $ from the order parameter $\Delta $ of
the bulk superconductor. This generates the paramagnetic instability of the
Andreev states, and the density of states of the NS structure exhibits a 
\textbf{\textit{single }}peak near zero energy. The theory developed in Ref. 
\cite{6a} essentially relies on the assumption of the repulsive electron
interaction in the normal region.

Maki and Haas \cite{7a} made the assumption that\textbf{\ }in noble metals
(Ag, Au) p-wave superconductors may occur with a transition temperature of
order $10$ mK. Below $T_{c}$ p -wave triplet superconductivity emerges
around the periphery of the cylinder. The diamagnetic current flowing in the
periphery is compensated by a quantized paramagnetic current in the opposite
direction thus providing a simple explanation for the reentrant effect. 
\begin{figure}[tbp]
\includegraphics[width=4in]{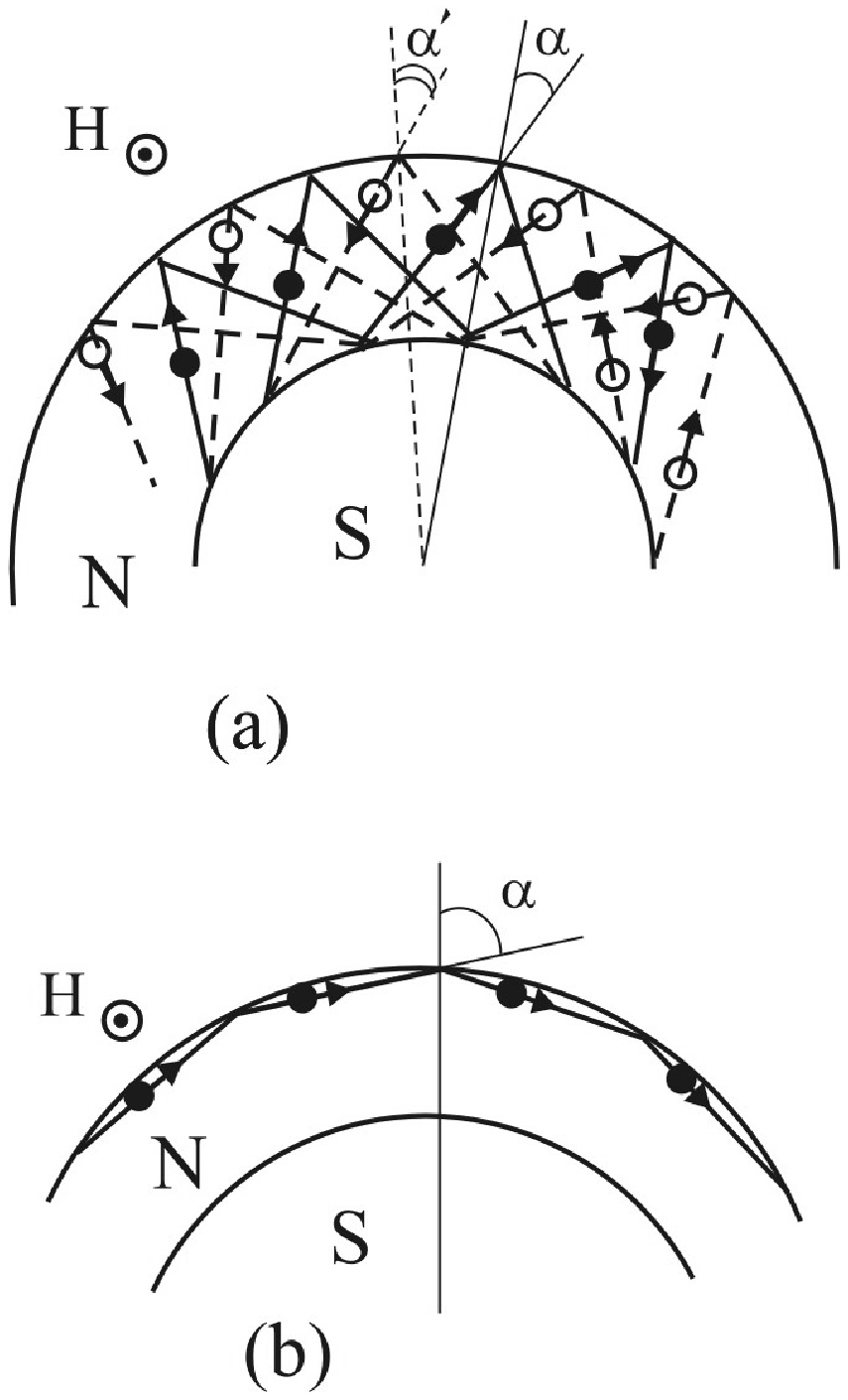}
\caption{Two classes of trajectories in the normal metal of the NS structure
in a magnetic field: trajectories forming the Andreev levels (a):
trajectories colliding only with the dielectric boundary (b ). This figure
has been taken from \protect\cite{9a}.}
\end{figure}
As in \cite{5a}, the authors of \cite{7a} also allow for paramagnetic
current in the system, which flows in the opposite direction to the
diamagnetic current. Its amplitude is sufficient to explain the reentrant
effect, but this theory says nothing about the temperature and field
dependences of magnetic susceptibility at ultra-low temperatures and in low
magnetic fields.

A theoretical basis for understanding the paramagnetic reentrant effect has
been proposed in \cite{8a,9a,10a}. The theory is essentially based on the
properties of the quantized levels of the NS structure. The Meissner effect
is rather special in a superconducting cylinder coated with a pure
normal-metal layer. The applied magnetic field generates superconducting
current in the surface layer whose thickness is equal to the field
penetration depth $\delta $. Simultaneously, the Aharonov-Bohm effect
generates persistent current (through the mechanism of the Andreev
scattering of quasiparticles) in the normal layer near the NS boundary. If N
and S metals are separated by a dielectric layer destroying the Andreev
scattering, the additional current disappears, and the Meissner effect
returns to its usual form. The levels with energies no more than $\Delta $ (2%
$\Delta $ is the gap of the superconductor) appear inside the normal metal
bounded by the dielectric (vacuum) on one side and the superconductor on the
other side. Because of the Aharonov-Bohm effect \cite{11a}, the spectrum of
the NS structure in a weak field is a function of the magnetic flux. The
specific feature of the Andreev quantum levels of the structure is that by
varying field $\mathrm{H}$ (or temperature $T$) each level in the well
periodically comes into coincidence with the chemical potential of the
metal. As a result, the state of the system suffer strong degeneracy, and
the density of states on the energy of the NS sample experiences resonance
spikes \cite{9a}. This contributes significantly to the magnetic moment and
causes a reentrant effect. Note that in \cite{9a} the calculation was
performed for orbital susceptibility. In \cite{6a} the explanation involves
the spin (Pauli) susceptibility of the system.

In this study we calculated the free energy of the NS structure and its
magnetic moment (current of magnetization) in the magnetic field. An
approximate analytical calculation was supplemented with a numerical one
based on the exact spectrum of Andreev levels \cite{12a} in the NS contact.
Our approach is not based on application of the Eilenberger equation. We
calculate the thermodynamic potential to obtain the magnetic moment.

\section{Theory and results}

\subsection{Spectrum of Quasiparticles of the NS Structure}

Consider a superconducting cylinder with the radius $R$ which is
concentrically embedded with a thin layer $d$ of a pure normal metal. The
structure is placed in a magnetic field $\vec{H}\left( 0,0,\mathrm{H}\right) 
$ oriented along the symmetry axis of the structure. It is assumed that the
field is weak to the extent that the effect of twisting of quasiparticle
trajectories becomes negligible. It actually reduces to the Aharonov-Bohm
effect \cite{11a}, i.e., allows for the increment in the phase of the wave
function of the quasiparticle moving along its trajectory in the vector
potential field.

We proceed with a simplified model of NS structure in which the order
parameter magnitude changes stepwise at the NS boundary ( $\Delta \left(
x\right) =\Theta \left( -x\right) $, $\mathrm{Im}\left( \Delta \right) =0$,
a bulk superconductor in the region $x<0$ and a normal metal layer in the
region $0\leq x\leq d$ ). It is also assumed that the magnetic field does
not penetrate into the superconductor. The coherent properties observed in
the pure normal metal can be attributed to its large "coherence" length $\xi
_{N}\left( T\right) =\frac{\hbar \cdot \mathrm{v}_{F}}{\pi \cdot k_{B}T}$ ( $%
\mathrm{v}_{F}$ is the Fermi velocity, $k_{B}$ is the Boltzmann constant) at
very low temperatures ($\xi _{N}\left( T\right) $ $\gg $ $d$). Besides, the
spectrum of quasiparticles was obtained assuming a negligible curvature of
the NS boundary.

One can easily distinguish two classes of trajectories, inside the normal
metal. One of them includes the trajectories which collide in succession
with the dielectric and NS boundaries (\textbf{Fig.1}). The quasiparticles
moving along these trajectories have energies $\left\vert E\right\vert
<\Delta $ and are localized inside the potential well bounded by a high
dielectric barrier ( $\simeq $ $1eV$ ) on one side and by the
superconducting gap $\Delta $ on the other side ($\Delta $ $=$ $3.56\cdot $ $%
k_{B}T_{c}/2$, $\Delta (\mathrm{Nb})$ $\approx $ $1.42$ meV). On its
collisions, the quasiparticle is reflected specularly from the dielectric
and experiences the Andreev scattering at the NS boundary \cite{12a}. We
introduce an angle $\alpha $ at which the quasiparticle hits the dielectric
boundary. The angle is measured in the positive direction from the normal to
the boundary (\textbf{Fig.1}). In this case the first class contains the
trajectories with $\alpha $ varying within the range $-\alpha _{c}\leq
\alpha \leq \alpha _{c}$ ( $\alpha _{c}$ is the angle at which the
trajectory touches the NS boundary, sin$\left( \alpha _{c}\right) $ $=$ $%
\frac{R}{R+d}$ ). 
\begin{figure}[tbp]
\includegraphics[width=4in]{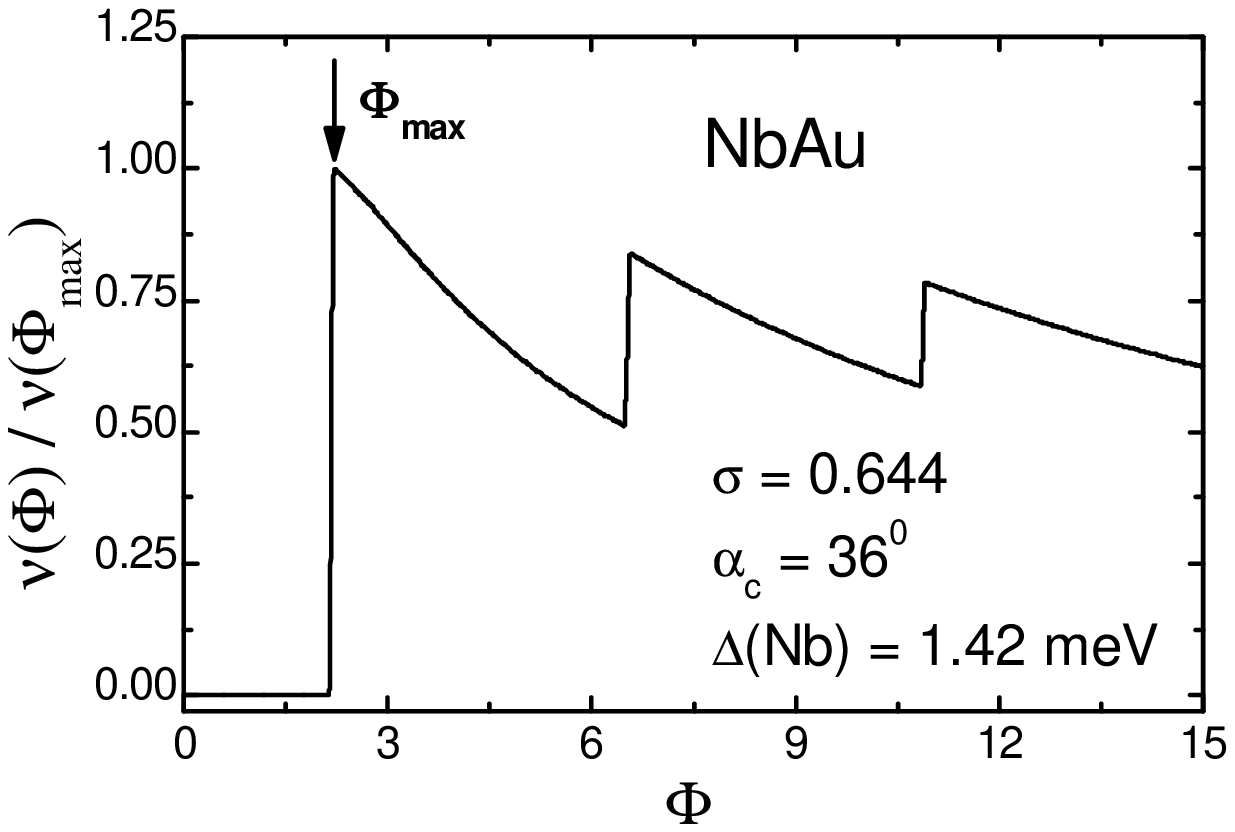}
\caption{The dependence of the density of states of the NS structure on the
magnetic flux $\Phi $ ( $E=E_{F}=0$ ). Normalization was performed for the
flux $\Phi _{{\protect\small max}}$ corresponding to the highest value of $%
\protect\nu \left( \Phi \right) $ ($\ \Phi _{max}\approx 2.175$ ).}
\end{figure}
Another class includes the trajectories whose spectra are formed by
collisions with the dielectric only, i.e., the trajectories with $\alpha
>\alpha _{c}$. The two groups of trajectories produce significantly
different spectra of quasiparticles. The distinctions are particularly
obvious in the presence of the magnetic field. The trajectories with $\alpha
\lesssim \alpha _{c}$ form a spectrum of Andreev levels which contains an
integral of the vector potential field. The spectrum characterizes the
magnetic flux through the area of the triangle between the quasiparticle
trajectory and the part of the NS boundary. It also determines the magnitude
of the screening current produced by "particles" and "holes" in the N layer.
These states are responsible for the reentrant effect. The trajectories with 
$\alpha >\alpha _{c}$ do not collide with the NS boundary. The states
induced by these trajectories are practically similar to the "whispering
gallery" type of states appearing in the cross section of a solid normal
cylinder in a weak magnetic field \cite{13a}, \cite{14a}. The size of the
caustic of these trajectories is of the order of the cylinder radius, i.e.,
they correspond in high magnetic quantum numbers. The spectrum thus formed
carries no information about the parameters of the superconductor, and it is
impossible to meet the resonance condition in this case. These states make a
paramagnetic contribution in the thermodynamics of the NS structure but
their amplitude is small ( $\sim 1/\left( k_{F}\cdot R\right) $ ). It is
therefore discarded from further consideration. Our interest will be
concentrated on the trajectories with $\left\vert \alpha \right\vert \leq
\alpha _{c}$.

The spectrum of quasiparticles of the NS structure can be obtained easily
using the multidimensional quasiclassical method generalized for the case of
the Andreev scattering in the system \cite{15a}, \cite{16a}. After collision
with the NS boundary the "particle" transforms into a "hole". The "hole"
travels practically along the path of the "particle" but in the reverse
direction (\textbf{Fig.1}).

The spectrum was derived by quantizing the adiabatic invariant $\frac{1}{%
2\pi }\oint \vec{P}\cdot d\vec{s}$ , where $\vec{P}$ $=$ $\vec{p}+$ $\frac{e%
}{c}\vec{A}$, $\vec{A}$ $=$ $\left( 0,A_{y}\left( x\right) ,0\right) $, $%
\vec{P}_{0}$ $=$ $\vec{p}_{0}-$ $\frac{\left\vert e\right\vert }{c}\vec{A}$
for a "particle" and $\vec{P}_{1}$ $=$ $\vec{p}_{1}+$ $\frac{\left\vert
e\right\vert }{c}\vec{A}$ for a "hole". Note that each collision with the NS
boundary multiplies the wave function amplitude of the quasiparticle by a
factor of $\exp \left( -i\arccos \left( {E/\Delta }\right) \right) $. 
\begin{figure}[tbp]
\includegraphics[width=4in]{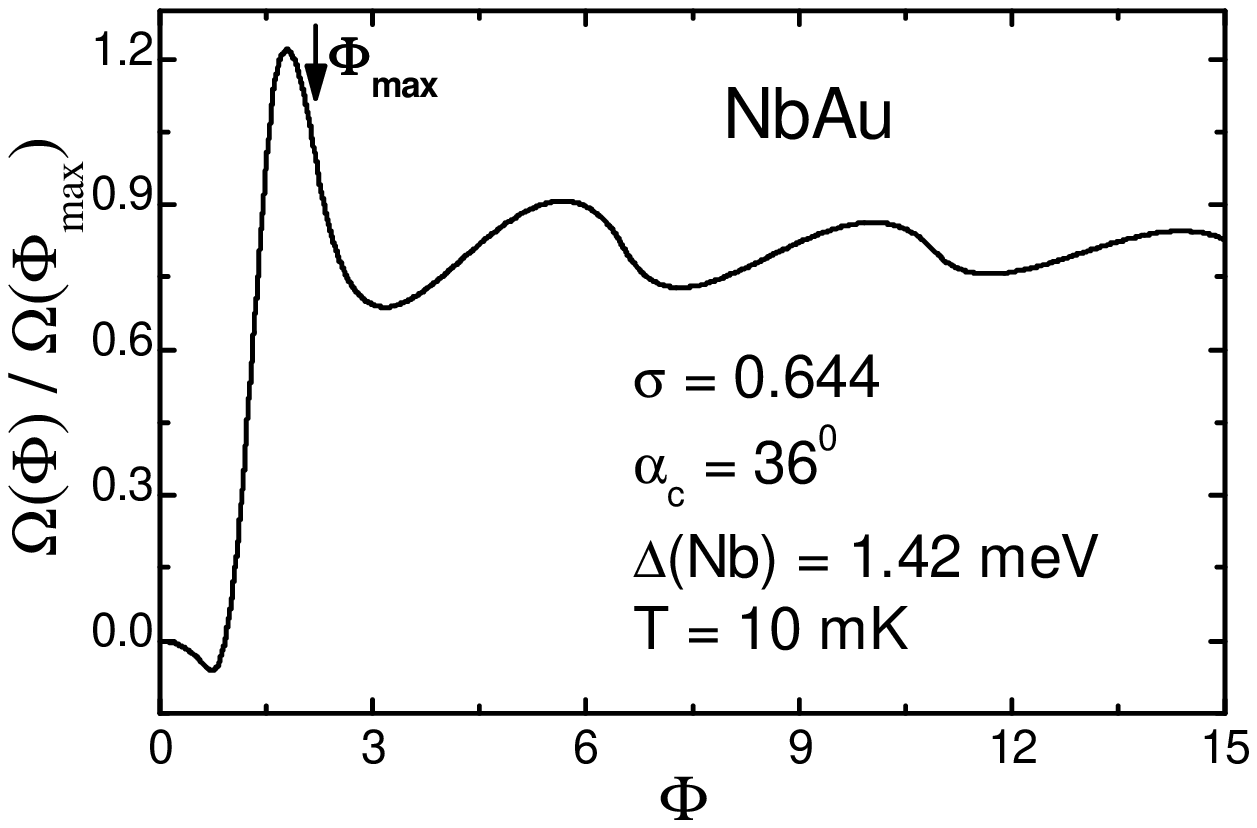}
\caption{Free energy (normalized per value $\Omega \left( \Phi _{\max
}\right) $) as a function of the flux $\Phi $.}
\end{figure}
Let $\mathrm{{\mathcal{L}}}_{0}$ be the length of the quasiparticle
trajectory between the collisions at the boundaries of the N layer. We thus 
\cite{8a}, \cite{9a} arrive at the expression for the spectrum of the
Andreev levels in the NS structure:%
\begin{widetext}
\begin{equation}
E_{n}\left( q,\alpha ,\Phi \right) =\frac{\pi \hbar \mathrm{v}_{L}(q)}{%
\mathrm{{\mathcal{L}}}_{0}}\left[ n+\frac{1}{\pi }\arccos \left( \frac{%
E_{n}\left( q,\alpha ,\Phi \right) }{\Delta }\right) -\frac{\tan \left(
\alpha \right) }{\pi }\Phi \right].   \label{E1}
\end{equation}

\end{widetext}Here $\mathrm{v}_{L}(q)={\sqrt{p_{F}^{2}-q^{2}}/m^{\ast }}$, $%
\mathrm{{\mathcal{L}}}_{0}$ is length of the quasiparticle trajectory, $p_{F}
$ is the Fermi momentum, $q$ is the quasiparticle momentum component along
the cylinder axis ( $\left\vert q\right\vert \leq p_{F}$ ), ${m^{\ast }}$ is
the effective mass of the quasiparticle, and $\Phi _{0}={hc/2e}$ is the
superconducting flux quantum. The factor $\Phi $ appearing in the last term
in Eq. (\ref{E1}) has the meaning of "phase"%
\begin{equation}
\Phi =\frac{2\pi }{\Phi _{0}}\int\limits_{0}^{d}A_{y}\left( x\right) dx
\label{E2}
\end{equation}%
which is dependent on the vector potential field $\vec{A}$ $=$ $\left(
0,A_{y}\left( x\right) ,0\right) $. The spectrum of Eq. (\ref{E1}) is
similar to Kulik's spectrum \cite{17a} for the current state of the SNS
contact. However, Eq. (\ref{E1}) includes an angle-dependent magnetic flux
instead of the phase difference of the contacting superconductors.

The length of the quasiparticle trajectory ( $2AB$ ) is readily found from 
\textbf{Fig.1} using the sine and cosine theorems:%
\begin{equation}
AB=d\cdot \left( \frac{\cos \left( \alpha \right) -\sqrt{\sin \left( \alpha
_{c}\right) ^{2}-\sin \left( \alpha \right) ^{2}}}{1-\sin \left( \alpha
_{c}\right) }\right)  \label{E3}
\end{equation}%
where sin$\left( \alpha _{c}\right) $ $=$ $\frac{R}{R+d}$, $-\alpha _{c}\leq
\alpha \leq \alpha _{c}$ . The spectrum in Eq. (\ref{E1}) was derived
assuming that the mean free path of the quasiparticles was much longer than
the cross-section perimeter of the cylinder and the requirement $d\ll R$ was
obeyed. In this limit $\mathrm{{\mathcal{L}}}_{0}$ $=$ $2AB$ $\cong $ ${%
2d/\cos \left( \alpha \right) }$ $\left( \underset{\alpha _{c}\rightarrow
\pi /2}{\lim }AB={d/\cos \left( \alpha \right) }\right) $ , i.e., the radius 
$R$ drops out from the expression for the spectrum. Although the boundary
curvature of the sample is disregarded, the information about its
cylindrical geometry is retained through a correct choice of the limits of
integration for the angle $\alpha $: $-\alpha _{c}\leq \alpha \leq \alpha
_{c}$. Putting $\mathrm{{\mathcal{L}}}_{0}$ $=$ ${2d/\cos \left( \alpha
\right) }$, we obtain the following expression for the spectrum ( as in \cite%
{9a} ):%
\begin{widetext}
\begin{equation}
E_{n}\left( q,\alpha ,\Phi \right) =\frac{\pi \hbar \mathrm{v}_{L}(q)\cos
\left( \alpha \right) }{2d}\left[ n+\frac{1}{\pi }\arccos \left( \frac{%
E_{n}\left( q,\alpha ,\Phi \right) }{\Delta }\right) -\frac{\tan \left(
\alpha \right) }{\pi }\Phi \right].  \label{E4}
\end{equation}
\end{widetext}\qquad The spectrum in Eq. (\ref{E4}) has an important feature 
\cite{8a}, \cite{9a}. As the "phase" $\Phi $ Eq. (\ref{E2})\ changes, the
density of states exhibits resonance spikes. Every time when the Andreev
level coincides with the chemical potential of the metal, the state of the
NS structure suffers strong degeneracy showing up spikes. The dependence of
the density of states upon the magnetic flux calculated numerically for the
NS system is illustrated in \textbf{Fig.2}.

Note that in \cite{1a}, \cite{18a} the diamagnetic current of NS structure
was calculated using $\alpha _{c}=\pi /2$ instead of the upper limit of
integration for the angle ( $\alpha _{c}<\pi /2$ ) and assuming implicitly
an infinitely large number of Andreev levels. Therefore, these results
cannot be conformed with the experimental findings. The reason is not only
that the calculation was made for a flat geometry rather than for a curved
NS boundary. Numerical analysis shows that an adequate interpretation of the
experimental magnetic moment-field dependence is possible only with a proper
choice of the upper limit of integration with respect to $\alpha $. If $%
\alpha >\alpha _{c}$ or $\alpha _{c}=\pi /2$, the consideration includes
effectively the states unrelated to the Andreev levels.

\subsection{Self-consistent equation}

To calculate the "phase" $\Phi \left( T,\mathrm{H}\right) $ from Eq. (\ref%
{E2}), we should know the distribution of the vector potential field inside
the normal metal. Zaikin \cite{1a} has shown that the proximity effect
caused the Meissner effect leads to an inhomogeneous distribution of the
vector potential field over the N layer of the structure:%
\begin{equation}
A_{y}\left( x\right) =\mu _{0}\mathrm{H}x+\mu _{0}j\cdot x\left( d-{x/2}%
\right)  \label{ins3}
\end{equation}%
where $\mu _{0}$\ is the permeability of free space ( the SI system of units
is employed, the geometry of the proximity model system is the same as in 
\cite{19ains} Fig 1). This expression can be obtained from the Maxwell
equation $rot\left( rot\left( \vec{A}\right) \right) $ $=$ $\mu _{0}\vec{j}$ 
$=$ $\left( 0,\mu _{0}j,0\right) $ assuming that the current density is
uniform over the cross-section of the conductor and the boundary condition $%
\left. \vec{A}\right\vert _{x=0}$ $=$ $0$, $\left. rot\left( \vec{A}\right)
\right\vert _{x=d}$ $=$ $\vec{B}\left( 0,0,\mu _{0}\mathrm{H}\right) $ is
met. The fact that the current density is constant in the N-layer follows
from spatial homogeneity of the density of Andreev levels over the whole
thickness of the N-layer. In cylindrical geometries, if the N-layer
thickness is not thin compared to the radius ( $d\gtrsim R$\ ), the current
density is not constant in space \cite{19a}.

The magnetic moment per unit length of the N-layer $M\left( \mathrm{H}%
\right) =-\frac{1}{\mu _{0}}\frac{d\Omega \left( T,\Phi \left( \mathrm{H}%
\right) \right) }{d\mathrm{H}}$ ( z-component ) and the current density $%
\vec{j}$ are related via a relation%
\begin{equation}
M\left( T,\mathrm{H}\right) =\frac{1}{2}\int\limits_{V_{N}}\left[ \vec{r}%
\times \vec{j}\left( \vec{r}\right) \right] _{z}dV=\frac{-1}{\mu _{0}}\frac{%
d\Omega }{d\mathrm{H}}  \label{ins1}
\end{equation}%
where $V_{N}$ is a volume of the N-layer unity height, $\Omega \left( T,\Phi
\right) $ is the free energy per unit length. Then, the current appears from
Eq. (\ref{ins1}) is a function of the magnetic flux $\Phi $ and temperature:%
\begin{equation}
j=-\frac{1}{\pi R^{2}d\cdot \mu _{0}}\frac{d\Omega \left( T,\Phi \left( 
\mathrm{H}\right) \right) }{d\mathrm{H}}  \label{ins4}
\end{equation}%
We can write down the self-consistent equation for $\Phi \left( T,\mathrm{H}%
\right) $ using Eqs. (\ref{E2}), (\ref{ins3}) and (\ref{ins4}):%
\begin{equation}
\Phi \left( T,\mathrm{h}\right) =\mathrm{h}+\eta \cdot M^{\ast }\left( \Phi
\right) \frac{\partial \Phi \left( T,\mathrm{h}\right) }{\partial \mathrm{h}}
\label{E5}
\end{equation}%
where $\mathrm{h}=\frac{\mathrm{H}}{\mathrm{H}_{0}}$, $\mathrm{H}_{0}=\frac{%
\Phi _{0}}{\pi d^{2}\cdot \mu _{0}}$, $\eta =\frac{d^{2}}{3R^{2}\Phi _{0}%
\mathrm{H}_{0}}$, $M^{\ast }\left( \Phi \right) =-\frac{d\Omega }{d\Phi }$ 
\cite{20FNT}. To describe the field effect on the magnetic moment%
\begin{equation}
M\left( T,\mathrm{H}\right) =\frac{M^{\ast }\left( T,\Phi \right) }{\mu _{0}%
\mathrm{H}_{0}}\cdot \frac{\partial \Phi \left( T,\mathrm{h}\right) }{%
\partial \mathrm{h}}  \label{ins2}
\end{equation}%
of a NS structure, it is necessary to find the dependence $\Phi \left( T,%
\mathrm{h}\right) $ from Eq. (\ref{E5}). After calculating the free energy $%
\Omega \left( T,\Phi \right) $ from the spectrum of Eq. (\ref{E4}), we can
estimate the magnetic moment Eq. (\ref{ins2}) using a solution of the
differential equation (\ref{E5}): $M\left( T,\mathrm{H}\right) $ $=$ $\frac{1%
}{\eta \cdot \mu _{0}\mathrm{H}_{0}}$ $\left( \Phi \left( T,\mathrm{h}%
\right) -\mathrm{h}\right) $. In this article, we used the "thermodynamic"
approach Eq. (\ref{ins4}) which leads to the first order differential
equation (\ref{E5}) for the function $\Phi \left( T,\mathrm{h}\right) $. 
\begin{figure}[tbp]
\includegraphics[width=4in]{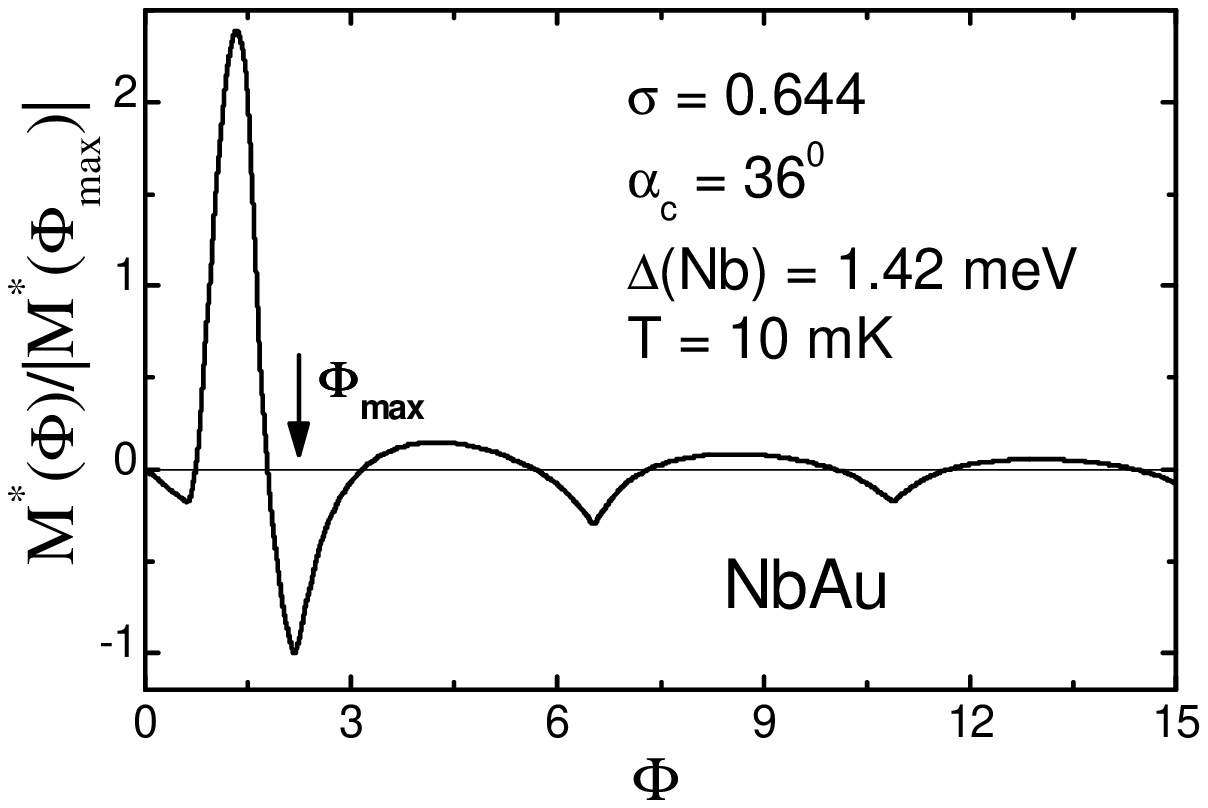}
\caption{The magnitude of $M^{\ast }\left( \Phi \right) /\left\vert M^{\ast
}\left( \Phi _{\max }\right) \right\vert $ as a function of the flux $\Phi $%
. }
\end{figure}
However, another approach based on of the Eilenberger formalism ( \cite{1a}, 
\cite{18a} ) yields an \textbf{algebraic} self-consistent equation ( Eq. 22
in (\cite{18a})\ ) for the "phase" $\Phi \left( T,\mathrm{h}\right) $:%
\begin{equation}
\Phi \left( T,\mathrm{h}\right) =\mathrm{h}+const\cdot j\left( \Phi \right) 
\label{ins6}
\end{equation}%
( in notation Eq., (\ref{E5}) ). In this equation the function $j\left( \Phi
\right) $ is described by the expression Eq., 13 in \cite{1a}. Clearly, both
approaches (\ref{E5}), (\ref{ins6}) will lead to quite different
dependencies of $M\left( T,\mathrm{H}\right) $ on magnetic field and
temperature. In our point of view, the self-consistent equation (\ref{ins6})
cannot be applied to the cylindrical NS structures. While deriving the
expression for a current \cite{1a}, the author assumed that $\alpha _{c}=\pi
/2$ (the case of a plane). This means account for the contribution
non-Andreev states ( $\alpha >\alpha _{c}$). For the calculation of
thermodynamic potential in equation (\ref{E5}) we shall use the actual
magnitude of the parameter $\alpha _{c}$ ( $sin(\alpha _{c})=R/(R+d)$ ). The
approximation $d<<R$ was used by us in the derivation of the spectrum (\ref%
{E4}) only. In other words, the neglect of the curvature of cylindrical
samples (i.e. the path length of quasiparticles was choosen as ${d/\cos
\left( \alpha \right) }$ (\ref{E3})) does not entail the need of account for
the states with $\alpha >\alpha _{c}$ for the cylindrical NS of structures (%
\textbf{Fig.1}).

\subsection{Analytical estimation of the magnetic moment of the NS structure}

We proceed from the expression for the free energy of a NS contact:%
\begin{equation}
\Omega =-k_{B}T\sum\limits_{n,q,\alpha ,s}\ln \left( 1+e^{-\frac{E_{n}\left(
q,\alpha \right) }{k_{B}T}}\right)  \label{E6}
\end{equation}%
where the summation is over the spin variable $s=\pm 1$ and all the states
related to the quasiparticles trajectories with $\left\vert \alpha
\right\vert \leq \alpha _{c}$ in Eq. (\ref{E4}). Then, we obtain the
following expression for the free energy per unit length $L$ 
\begin{widetext}
\begin{equation}
\Omega \left( \Phi \right) =-\frac{R\cdot k_{B}T}{\pi \hbar ^{2}}%
\sum\limits_{n=-\infty }^{n=\infty }\int_{-\alpha c}^{\alpha
_{c}}\int_{-p_{F}}^{p_{F}}\ln \left( 1+\exp \left( -\frac{E_{n}\left(
q,\alpha ,\Phi \right) }{k_{B}T}\right) \right) \sqrt{p_{F}^{2}-q^{2}}\cos
\left( \alpha \right) dqd\alpha   \label{E15}
\end{equation}

\end{widetext}where the energy $E_{n}\left( q,\alpha ,\Phi \right) $ is
given by the exact expression for the spectrum in Eq. (\ref{E4}). For
simplification, we introduce the dimensionless quantities $\varepsilon _{n}=%
\frac{E_{n}\left( q,\alpha ,\Phi \right) }{\Delta }$, $\sigma =\frac{\hbar
\cdot p_{F}}{2d\cdot \Delta \cdot m^{\ast }}$, $-1\leq \varepsilon _{n}\leq
1 $, ( $\Delta $ is the superconducting gap ) and perform the change of
variables $\left\{ q,\alpha \right\} \rightarrow \left\{ u,v\right\} $:%
\begin{equation}
\left\{ 
\begin{array}{c}
u={\sigma \cdot \sqrt{1-\left( \frac{q}{p_{F}}\right) ^{2}}\cdot \cos \left(
\alpha \right) } \\ 
v={\sigma \cdot \sqrt{1-\left( \frac{q}{p_{F}}\right) ^{2}}\cdot \sin \left(
\alpha \right) }%
\end{array}%
\right. .  \label{E16}
\end{equation}%
The spectrum and the free energy become%
\begin{equation}
\varepsilon _{n}=\left[ n\pi +\arccos \left( \varepsilon _{n}\right) \right]
\cdot u-\Phi \cdot v\text{,}  \label{E17}
\end{equation}%
\begin{equation}
\Omega \left( \Phi \right) =c_{1}T\sum\limits_{n=0}^{n=\infty
}\iint\limits_{S}\frac{u\ln \left( 2\cosh \left( \frac{\varepsilon _{n}}{%
2c_{2}T}\right) \right) dudv}{\sqrt{\sigma ^{2}-u^{2}-v^{2}}}  \label{E18}
\end{equation}%
where $c_{1}=-2\frac{R\cdot \Delta \cdot c_{2}}{\pi }\cdot \left( \frac{p_{F}%
}{\sigma \cdot \hbar }\right) ^{2}$, $c_{2}=\frac{k_{B}}{\Delta }$, $0\leq
u\leq \sigma $, $-\sigma \sin \left( \alpha _{c}\right) \leq v\leq \sigma
\sin \left( \alpha _{c}\right) $, $\varepsilon _{n}\overset{def}{=}%
\varepsilon _{n}\left( u,v,\Phi \right) $, an integration domain $S$ is a
sector of a circle of radius $\sigma $. In the expression (\ref{E18}) we
also took into account the symmetry of the spectrum in Eq. (\ref{E17}):%
\begin{equation}
\varepsilon _{-\left\vert n\right\vert }\left( u,v,\Phi \right)
=-\varepsilon _{\left\vert n\right\vert -1}\left( u,-v,\Phi \right) \text{.}
\label{E8}
\end{equation}%
Making use of the relation%
\begin{equation}
\frac{d\varepsilon _{n}}{d\Phi }=-\frac{v\cdot \sqrt{1-\varepsilon _{n}^{2}}%
}{u+\sqrt{1-\varepsilon _{n}^{2}}}  \label{ins5}
\end{equation}%
we evaluate the derivative of the free energy with respect to the flux $%
M^{\ast }\left( \Phi \right) =-\frac{d\Omega }{d\Phi }$:%
\begin{equation}
M^{\ast }\left( \Phi \right) =c_{3}\sum\limits_{n=0}^{n=\infty
}\iint\limits_{S}\frac{uv\tanh \left( \frac{\varepsilon _{n}}{2c_{2}\cdot T}%
\right) \sqrt{1-\varepsilon _{n}^{2}}dudv}{\left( u+\sqrt{1-\varepsilon
_{n}^{2}}\right) \sqrt{\sigma ^{2}-u^{2}-v^{2}}}  \label{E19}
\end{equation}%
where $c_{3}=\frac{R\cdot \Delta }{\pi }\left( \frac{p_{F}}{\sigma \cdot
\hbar }\right) ^{2}$. Eqs. (\ref{ins2}), (\ref{E5}) and (\ref{E19}) fully
determine the non-linear magnetic response of a cylindrical NS structure to
an externally applied magnetic field $\mathrm{H}$.

The integral expression of Eq. (\ref{E19}) suggests that $M^{\ast }\left(
\Phi \right) $ is the odd function of the flux $\Phi $: $M^{\ast }\left(
\Phi \right) $ $=$ $-M^{\ast }\left( -\Phi \right) $. A linear term of the
function $M^{\ast }\left( \Phi \right) $ has been determined from an
approximate estimation of the integral in Eq. (\ref{E19}). This calculation
is similar to that in the Attachment of \cite{9a}. The final expression for
the magnetic moment is%
\begin{equation}
M\left( T,\mathrm{h}\right) \simeq M_{0}\sum\limits_{n=0}^{n_{0}}\frac{\ln
\cosh \left( \frac{T_{A}\cdot \tilde{n}}{2T}\right) }{\tilde{n}^{3}\left[
1+\left( \frac{\Phi \left( T,\mathrm{h}\right) }{\pi \cdot \tilde{n}}\right)
^{2}\right] ^{\frac{{3}}{2}}}  \label{E14}
\end{equation}%
where $\tilde{n}$ $=$ $n+\frac{1}{2}$, $T_{A}$ $=$ $\frac{\hbar \cdot 
\mathrm{v}_{F}}{2\pi d\cdot k_{B}}$ is Andreev temperature, $M_{0}=$ $-$ $%
c_{3}$ $\cdot $ $\sigma ^{2}\cdot \left( \frac{T}{T_{A}}\right) $ $\cdot $ $%
\Phi \left( T,\mathrm{h}\right) \cdot \frac{\partial \Phi \left( T,\mathrm{h}%
\right) }{\partial \mathrm{h}}$, the "phase" $\Phi \left( T,\mathrm{h}%
\right) $ is a solution of the differential equation (\ref{E5}), $n_{0}$ is
the number of Andreev levels in the potential well ( $n_{0}\approx \frac{%
\Phi \left( T,\mathrm{h}\right) }{\pi }\tan \left( \alpha _{c}\right) $ ).
Eq. (\ref{E14}) shows that the magnetic moment is diamagnetic in the range
of small fields ( $\Phi \left( \mathrm{h}\right) =const\cdot \mathrm{h}$, $%
const>0$ ) and allows for the contributions of "particles" and "holes".

\subsection{Numerical results}

Let us compare two approaches described above for the calculation of the
magnetic moment of the NS structure. \textbf{Fig.5} shows the function $%
M^{\ast }\left( \Phi \right) $ and dependency of the current on the magnetic
flux obtained in the Green's function approach. For comparison, we obtained
the dependence $M^{\ast }\left( \Phi \right) $ at the same value $\alpha
_{c}=\pi /2$ as was used in the derivation of the formula $j\left( \Phi
\right) $ in \cite{18a}. In the initial part (linear in the $\Phi $) both
curves coincide. In this approximation ($\Phi \left( \mathrm{h}\right)
=const\cdot \mathrm{h}$) the self-consistent equation (\ref{ins2}) turns
into Eq. (\ref{ins6}). Thus, at small values of the magnetic field we\textbf{%
\ }would obtain\textbf{\ }the same field dependence of the magnetic moment $%
M\left( T,\mathrm{h}\right) $ for the NS structure in both approaches. 
\begin{figure}[tbp]
\includegraphics[width=4in]{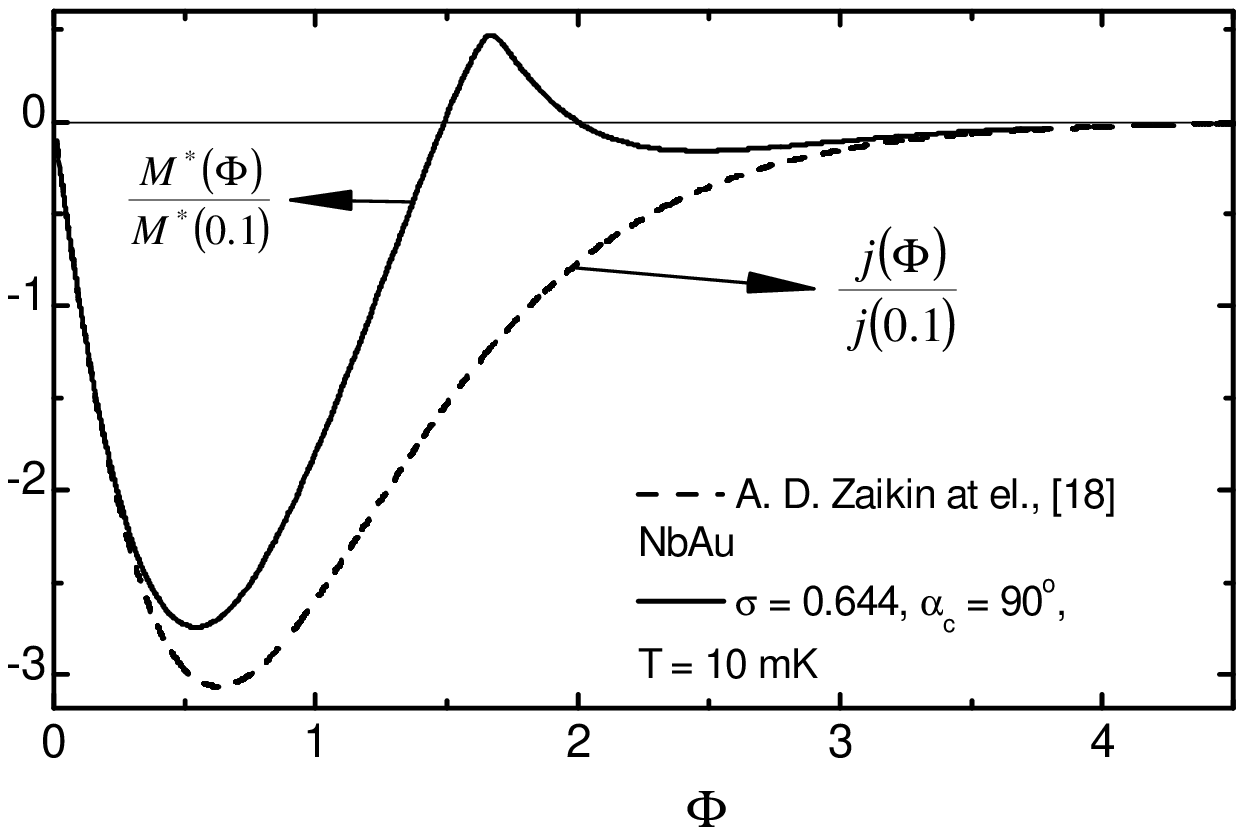}
\caption{The magnitudes $M^{\ast }\left( \Phi \right) /\left\vert M^{\ast
}\left( 0.1\right) \right\vert $ and $j\left( \Phi \right) /\left\vert
j\left( 0.1\right) \right\vert $ as the functions of the flux $\Phi $ (for
explanation see text).}
\end{figure}
However, in large fields the behavior $M\left( T,\mathrm{h}\right) $ is
quite different. To calculate $M\left( T,\mathrm{h}\right) $ from Eq. (\ref%
{ins2}), we have used the following physical values of the NS structure: $%
R=8.3$ $\mu $m, $d=3.2$ $\mu $m, ($\alpha _{c}=36^{0}$), $\mathrm{v}%
_{F}\left( \text{Au}\right) =1.4\cdot 10^{8}$ cm/s, $\Delta \left( \text{Nb}%
\right) =1.42$ meV ( $\sigma =0.644$, $\eta \cdot c_{3}=5.3\cdot 10^{3}$, $%
\mathrm{H}_{0}=51$A/m $=0.64$ Oe ). The selected parameters are close to
those used in the experiment \cite{3a,4a,20a}.

The results of calculation according to formulas (\ref{E18}) and (\ref{E19})
are illustrated in \textbf{Fig. 3, 4}. While plotting \textbf{Fig. 3,} the
nonzero quantity $\Omega \left( \Phi =0\right) $ was omitted. The dependence 
$M^{\ast }\left( \Phi \right) =-\frac{d\Omega }{d\Phi }$ (\textbf{Fig. 4})
crosses the abscissa thereby determining singular points of the differential
equation (\ref{E5}). The dependence $\Phi \left( \mathrm{h},\mathrm{T}%
\right) $ calculated through numerical solution of the self-consistency
equation (\ref{E5}) exhibits jumps and is illustrated in \textbf{Fig.6a} for
the branches corresponding to the minimum of the Gibbs free energy \cite%
{Gibbs}:%
\begin{equation}
\mathcal{G}\left( T,\mathrm{H}\right) =\Omega \left( T,\mathrm{H}\right) +%
\frac{1}{2\mu _{0}}\int\limits_{V_{N}}\left( \vec{B}-\mu _{0}\vec{H}\right)
^{2}dV  \label{Gibbs}
\end{equation}%
where $\vec{B}=rot\left( \vec{A}\right) $, $\vec{H}=\vec{H}\left( 0,0,%
\mathrm{H}\right) $. The magnetic moment $M\left( \mathrm{h}\right) $ and
the free energy $\Omega \left( \mathrm{h}\right) $ as functions of the
magnetic field are shown in \textbf{Fig. 7a, Fig. 6b.} Each jump $\Delta
\Omega $ of the free energy (see \textbf{Fig. 6b}) is accompanied by the
jump of the magnetic moment $\Delta M$ (see \textbf{Fig. 7a}) in such a way
that the Gibbs free energy (\ref{Gibbs}) is a continuous function of versus
the magnetic field $\mathrm{h}$. We have not performed an analysis of
behavior Gibbs free energy (\ref{Gibbs}) near the points where the magnetic
moment has jumps because it is beyond the semi-classical approximation
adopted in this article.

\section{Conclusions}

The goal of our study was to interpret the experiments performed by A. C.
Mota et al., \cite{3a,4a}, who detected an anomalous behavior of the
magnetic susceptibility of the NS structure in a weak magnetic field at
millikelvin temperatures. 
\begin{figure}[tbp]
\includegraphics[width=4in]{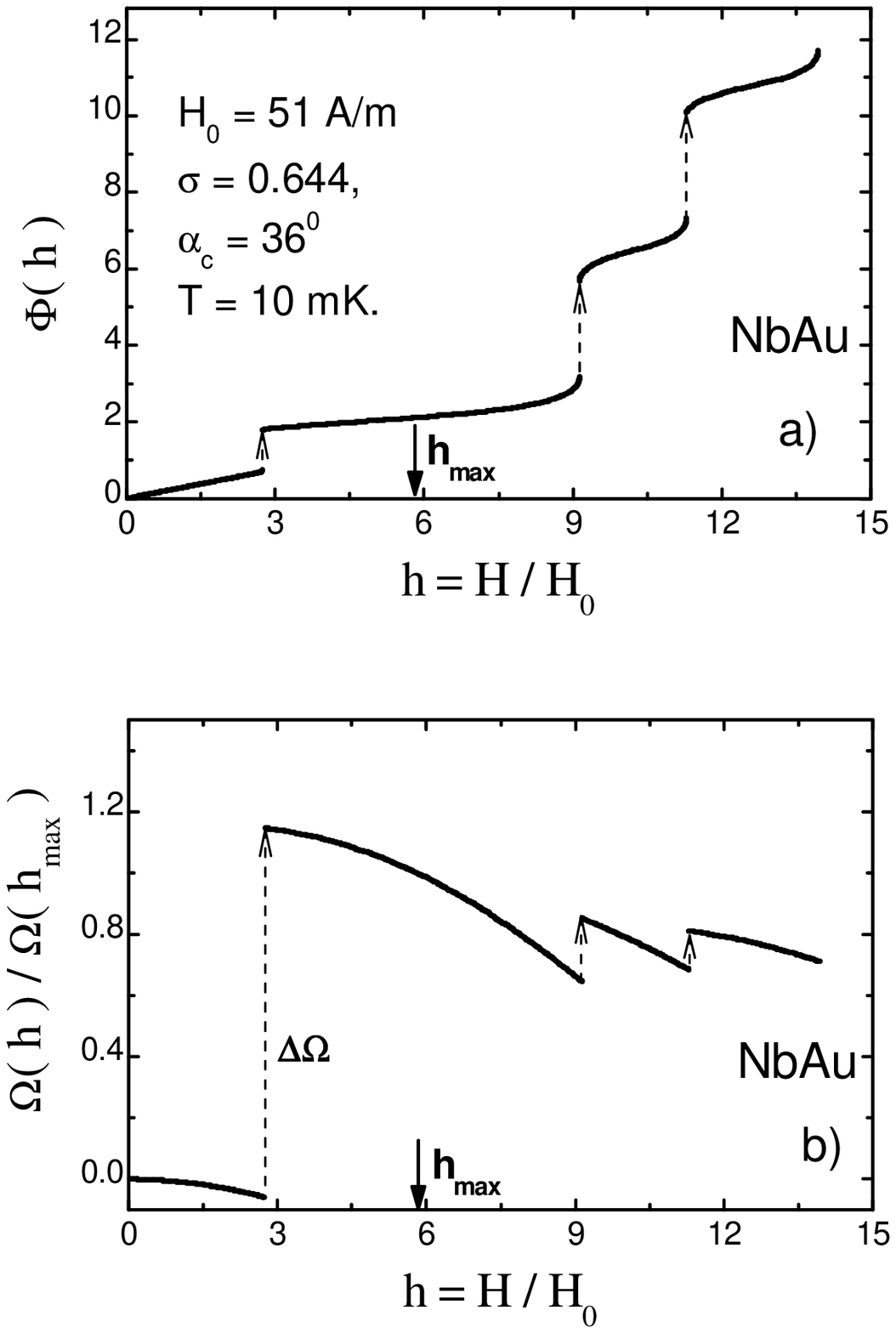}
\caption{a) the dependence of $\Phi \left( T,\mathrm{h}\right) $ on the
magnetic field $\mathrm{h}$. b) the relative free energy $\Omega \left( 
\mathrm{h}\right) /\Omega \left( \mathrm{h}_{\max }\right) $ ($\Phi \left( 
\mathrm{h}_{\max }\right) =\Phi _{\max }$) as a function of the magnetic
field $\mathrm{h}$.}
\end{figure}
Previously \cite{8a,9a}, the anomalous behavior of the NS structure was
attributed to the properties of the quantized Andreev levels depending on
the magnetic flux that varies with the temperature and magnetic field. We
used the "thermodynamic" approach for the calculation of the magnetic moment
of normal region of NS structure. Within the framework of the
self-consistent equation (\ref{E5}), we have managed to trace the role of
the parameter $\alpha _{c}$ in thermodynamics of NS of structures (\textbf{%
Fig. 4} and \textbf{Fig. 5}). Failure connected with the use of the
quasi-classical Green-function technique \cite{1a} for the explanation of
the experimental data (Mota et al) is a consequence of account for states
which are non-Andreev ones ( $\alpha >\alpha _{c}$ ) for the cylindrical NS
structures. Geometrically, this can be seen from figure \textbf{Fig. 1}. The
quasiparticle trajectories ( $\alpha >\alpha _{c}$ ) hitting the dielectric
boundary only are responsible for the paramagnetic current of a small
amplitude ( $\sim 1/(k_{F}\cdot R)$) (we neglected this current).

The proximity effect is crucial for the reentrant effect. The amplitude of
the resonance spikes in the density of states strongly depends on the
probability of the Andreev reflection at the NS boundary. It is therefore
assumed that the normal metal and the superconductor are in a good electric
contact. The spectrum Eq. (\ref{E4}) was obtained by the method of
multidimensional quasi-classical approach \cite{15a,16a}. 
\begin{figure}[tbp]
\includegraphics[width=4in]{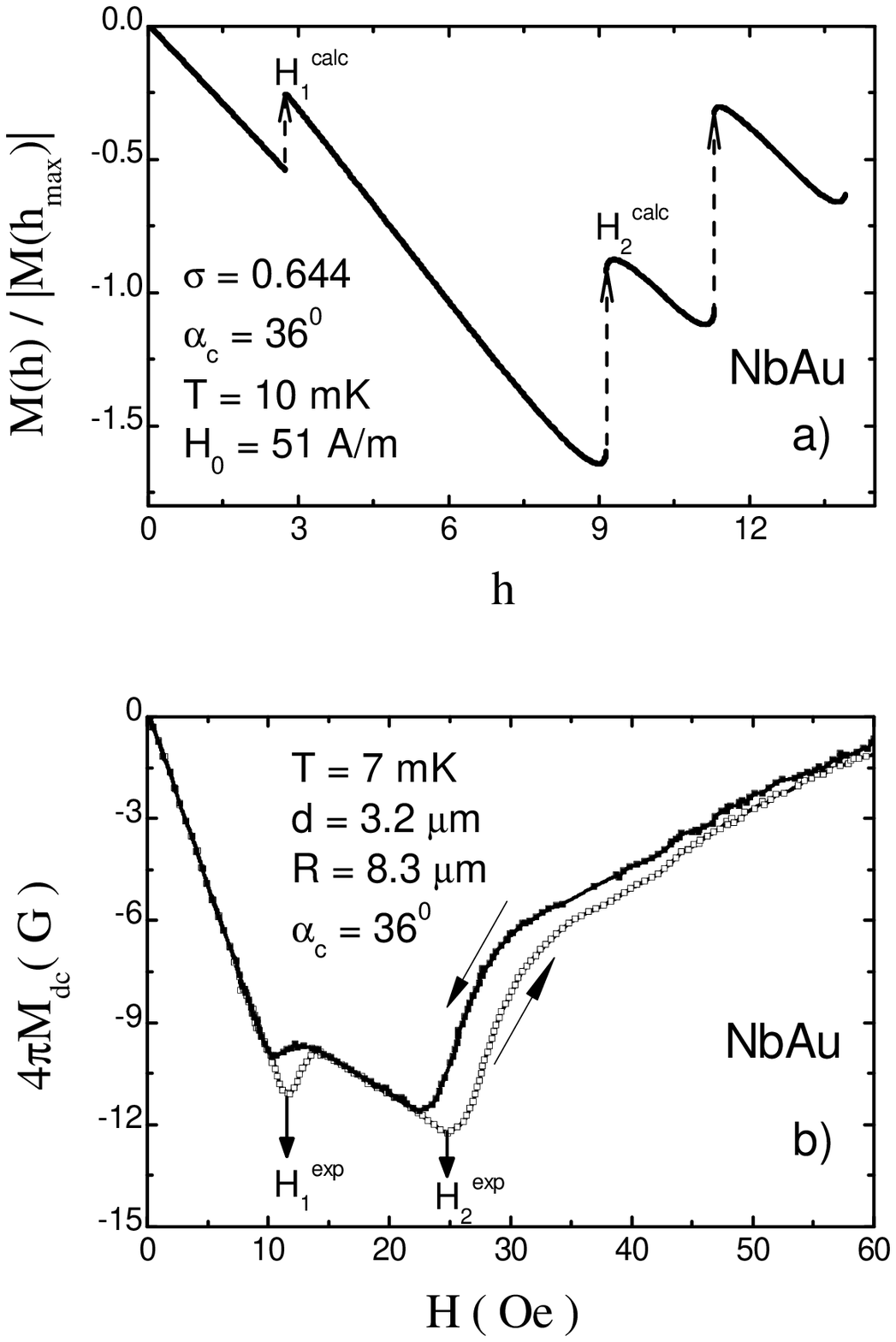}
\caption{a) The magnetic moment of the NS structure versus the magnetic
field $\mathrm{h}$. b) Isothermal dc-magnetization curve at $T=7$ mK for the
sample 41AuNb \protect\cite{20a}. Used by courtesy of A. C. Mota.}
\end{figure}
In doing so, we assumed (i) the condition of smallness of N-layer thickness
in comparison with the radius of the cylindrical superconductor, (ii) the
validity of the model of a stepwise-varing order parameter of the structure,
(iii) the independence of $\Delta $ on the magnetic field. This permitted us
to pass over from the curved NS boundary to a flat one. The information
about the cylindrical geometry of the sample was retained because the
critical angle at which a quasiparticle hits the dielectric boundary $\alpha 
$ is smaller than $\alpha _{c}$. The problem was further simplified by
assuming that the reflected quasiparticle performed a reciprocating motion,
i.e., a "particle" and a "hole" pass along the same trajectory but in
opposite direction. Actually, here there exists a lot of quasi reciprocating
trajectories with the energies near $\varepsilon \sim 0$. These trajectories
lead to the spikes in the density of states and were taken into account for
numerical computation. The numerical calculation shows that the nonlinearity
of flux $\Phi \left( \mathrm{T},\mathrm{h}\right) -$ field dependence ( $%
\mathrm{T=}$ constant ) (\textbf{Fig.6a}) gives rise to quite interesting
features of $M\left( T,\mathrm{H}\right) $. The magnetic moment in the N
layer appears to be always diamagnetic. The paramagnetic contribution to
current (the paramagnetic reentrant effect) was not detected. However, we
have obtained the stepwise change in absolute value of the magnetic moment
with increasing magnetic field (\textbf{Fig.7a}). This behavior can be
interpreted as an appearance in the magnetic moment the paramagnetic
additives. A behaviour of the NS structure changes from one stable state to
another and the magnetic field penetrates further into a bulk of the N
layer. The new state has smaller absolute value of the diamagnetic moment,
which is interpreted experimentally as an evidence of a paramagnetic
addition in the system (\textbf{Fig.7b}). When the field grows further, $%
\left\vert M\left( \mathrm{H}\right) \right\vert $ increases again until its
new value makes the system to jump to a next stable state with a smaller
absolute value of the diamagnetic moment, and the magnetic field penetrates
deeper inside the normal metal (\textbf{Fig.7a}). The number of the moment
jumps depends on the number of Andreev levels in the NS structure. Under the
isothermal condition, the values of the magnetic field at which jumps occur,
do not coincide when the magnetic field changes from small to larger values
and in the opposite direction because of a special dependence of the Gibbs
free energy on the field. This sort of hysteresis was observed
experimentally in \cite{4a,20a} (\textbf{Fig.7b}).

Numerical comparison between data presented at \textbf{Fig.7b} and \textbf{%
Fig.7a} shows that $M\left( T,\mathrm{H}\right) $ (\textbf{Fig. 7a}) gives
the qualitative description of the experimental data which obey the scaling
rule: $\mathrm{H}_{2}^{\exp }/\mathrm{H}_{1}^{\exp }\simeq \mathrm{H}%
_{2}^{calc}/\mathrm{H}_{1}^{calc}\simeq 5/2$. For the quantitative
description of the temperature and magnetic field dependence of the magnetic
moment NS of structures, it will be important to take into account the exact
spectrum of Andreev levels, the latter is supposedly possible within the
framework of the Bogoliubov -- de Gennes equations only.

Note that our consideration was entirely based on the model of free
electrons without account of strong electron-electron repulsion.

\section{Acknowledgement}

The authors thank A. N. Omelyanchouk, I. O. Kulik, I. V. Krive and G. I.
Japaridze for useful discussions.

\end{document}